\documentclass[useAMS,usenatbib]{mn2e}
\usepackage{aas_macros,graphicx,times,multirow,amsmath}
\usepackage[]{color}

\title[X-ray and radio observations of \src]{X-ray and radio observations of the magnetar \src\ and its dust-scattering halo}
\author[P. Esposito et al.]{P.~Esposito,$^{1}$\thanks{E-mail: paoloesp@iasf-milano.inaf.it}
 A.~Tiengo,$^{1,2,3}$ N.~Rea,$^{4}$ R.~Turolla,$^{5,6}$ A.~Fenzi,$^{1,7}$ A.~Giuliani,$^1$
\newauthor  G.~L.~Israel,$^{8}$ S.~Zane,$^6$ S.~Mereghetti,$^1$ A.~Possenti,$^{9}$ M.~Burgay,$^{9}$ L.~Stella,$^{8}$
\newauthor D.~G\"otz,$^{10}$ R.~Perna,$^{11}$ R.~P.~Mignani$^{6,12}$ and P.~Romano$^{13}$
\smallskip\\
$^1$INAF -- Istituto di Astrofisica Spaziale e Fisica Cosmica - Milano, via E. Bassini 15, I-20133 Milano, Italy\\
$^2$IUSS -- Istituto Universitario di Studi Superiori, piazza della Vittoria 15, I-27100 Pavia, Italy\\
$^3$INFN -- Istituto Nazionale di Fisica Nucleare, Sezione di Pavia, via A. Bassi 6, I-27100 Pavia, Italy\\
$^4$Institut de Ci\`encies de l'Espai (IEEC--CSIC), Campus UAB,  Torre C5, 2a planta, E-08193 Barcelona, Spain\\
$^5$Dipartimento di Fisica e Astronomia, Universit\`a di Padova, Via F. Marzolo 8, I-35131 Padova, Italy\\
$^6$Mullard Space Science Laboratory, University College London, Holmbury St. Mary, Dorking, Surrey, RH5 6NT, UK\\
$^7$Universit\`a degli Studi di Milano, Dipartimento di Fisica, via G. Celoria 16, I-20133 Milano, Italy\\
$^8$INAF -- Osservatorio Astronomico di Roma, via Frascati 33, I-00040 Monteporzio Catone, Italy\\
$^9$INAF -- Osservatorio Astronomico di Cagliari, loc. Poggio dei Pini, strada 54, I-09012 Capoterra, Italy\\
$^{10}$AIM Ð CEA/Irfu/Service dÕAstrophysique, Orme des Merisiers, F-91191 Gif-sur-Yvette, France\\
$^{11}$JILA and Department of Astrophysical and Planetary Sciences, University of Colorado, Boulder, CO 80309, USA\\
$^{12}$Institute of Astronomy, University of Zielona G\'ora, Lubuska 2, P-65265 Zielona G\'ora, Poland\\
$^{13}$INAF -- Istituto di Astrofisica Spaziale e Fisica Cosmica - Palermo, via U. La Malfa 153, I-90146 Palermo, Italy
}
\date{Accepted \ldots. Received \ldots; in
original form \ldots} \pagerange{\pageref{firstpage}--\pageref{lastpage}} \pubyear{2012}

\def\LaTeX{L\kern-.36em\raise.3ex\hbox{a}\kern-.15em
    T\kern-.1667em\lower.7ex\hbox{E}\kern-.125emX}

\def\xmm {\emph{XMM-Newton}}
\def\cxo {\emph{Chandra}}
\def\swift {\emph{Swift}}
\def\frm {\emph{Fermi}}

\def\xte {\emph{RXTE}}

\def\src {Swift\,J1834.9--0846}
\def\flux {\mbox{erg cm$^{-2}$ s$^{-1}$}}
\def\lum {\mbox{erg s$^{-1}$}}
\def\nh {$N_{\rm H}$}

\begin{document}

\label{firstpage}
\maketitle
\begin{abstract}
We present a long-term study of the 2011 outburst of the magnetar \src\ carried out using new \cxo\ observations, as well as all the available \swift, \xte, and \xmm\ data. The last observation was performed on 2011 November 12, about 100 days after the onset of the bursting activity that had led to the discovery of the source on 2011 August 07. This long time span enabled us to refine the rotational ephemeris and  observe a downturn in the decay of the X-ray flux. Assuming a broken power law for the long-term light curve, the break was at $\sim$46 d after the outburst onset, when the decay index changed from $\alpha\sim 0.4$ to $\sim$4.5. The flux decreased by a factor $\sim$2 in the first $\sim$50 d and then by a factor $\sim$40 until November 2011 (overall, by a factor $\sim$70 in $\sim$100 d). At the same time, the spectrum, which was well described by an absorbed blackbody all along the outburst, softened, the temperature dropping from  $\sim$1  to $\sim$0.6 keV. Diffuse X-ray emission extending up to 20$''$ from the source was clearly detected in all \cxo\ observations. Its spatial and spectral properties, as well as its time evolution, are consistent with a dust-scattering halo due to a single cloud located at a distance of $\approx$200 pc from \src, which should be in turn located at a distance of $\sim$5 kpc. Considering the time delay of the scattered photons, the same dust cloud might also be responsible for the more extended emission detected in \xmm\ data taken in September 2011. We searched for the radio signature of \src\ at radio frequencies using the Green Bank Radio Telescope and in archival data collected at Parkes from 1998 to 2003. No evidence for radio emission was found, down to a flux density of 0.05 mJy (at 2 GHz) during the outburst and $\sim$0.2-Ð0.3 mJy (at 1.4 GHz) in the older data.
\end{abstract}

\section{Introduction}
Anomalous X-ray pulsars (AXPs) and soft gamma-ray repeaters (SGRs)\footnote{See the McGill Pulsar Group SGR/AXP catalogue at the web page http://www.physics.mcgill.ca/~pulsar/magnetar/main.html.} are now generally accepted to be magnetars, i.e. neutron stars endowed with exceptionally strong magnetic fields (e.g. \citealt{mereghetti08}). According to the magnetar model \citep*{thompson95,thompson96,tlk02}, the decay of this magnetic field provides most of the energy that powers the emission of the neutron star. This scenario is supported by the fact that the surface dipole magnetic fields inferred under standard assumption for SGRs and AXPs from their rotational parameters are above, or at the high end of, those of ordinary pulsars. In fact they often exceed $10^{14}$ G,\footnote{For some magnetars the surface dipole magnetic field has been estimated also with other methods, always obtaining values of $\sim$$10^{14}$ G (e.g. \citealt{thompson95,thompson01,vietri07,israel08short}).} although for SGR\,0418+5729 an upper limit as low as $7.5\times10^{12}$ G has been reported \citep{esposito10short,rea10short,turolla11}.

Magnetars undergo unpredictable periods of bursting activity and enhanced luminosity, lasting from just days to months. The advent of large-field-of-view, sensitive X-ray monitors on board \swift\ and \frm\ proved to be very effective in catching SGR-like bursts, normally $\sim$0.1-s long and with luminosities of $10^{39}$--$10^{42}$ \lum\ --- bright enough to be detectable almost anywhere in the Galaxy. This has significantly contributed to the growth of the number of known magnetars (e.g. \citealt{rea11}). However, when dormant (often for decades), magnetars can remain well concealed among hundreds of other unidentified sources present in the X-ray catalogues or simply be too faint to be detected. It is therefore likely that a potentially large number of Galactic magnetars have yet to be discovered (e.g. \citealt{muno08}).

A good example of this is the magnetar catalogued as \src. On 2011 August 07 at 19:57:46.36 UT the Burst Alert Telescope (BAT; \citealt{barthelmy05short}) aboard \swift\  was triggered by a short SGR-like burst (trigger 00458907; \citealt{delia11short}) and  $\sim$330 s after the trigger the X-Ray Telescope (XRT; \citealt{burrows05short}) began observing the field of the burst. This led to the discovery of an X-ray source with a 2--10 keV flux of a few $10^{-11}$ \flux\ which was undetectable in previous observations of the field (with an upper limit on the flux of $\la$$10^{-15}$ \flux; \citealt{kargaltsev12short}). A few hours later, a second burst from the source direction was recorded by the \emph{Fermi}/GBM \citep{kargaltsev12short}, while a further burst triggered the \swift/BAT again on 2011 August 29 (at 23:41:12.04 UT, trigger 501752; \citealt{hoversten11short}). This, together with the discovery of pulsations at 2.48 s with the \emph{Rossi X-ray Timing Explorer} (\xte), confirmed the magnetar nature of the source \citep{gogus11}.

Subsequent observations allowed the determination of the spin-down rate of \src\ ($\sim$$8\times10^{-12}$ s s$^{-1}$; \citealt{kuiper11,kargaltsev12short}). The 2--10 keV  spectrum is very absorbed ($N_{\rm H}>10^{23}$ cm$^{-2}$) and can be modelled by either a steep power law ($\Gamma\sim4$) or a hot blackbody ($kT\sim1$ keV; \citealt{kargaltsev12short}). The decay of the X-ray flux for the first $\sim$50 days was consistent with a power law, $F\propto t^{-\alpha}$ with $\alpha\sim0.5$.

\src\ was surrounded by an X-ray nebula  with a complex spatial structure. The emission within 50$''$ had a symmetrical shape and has been interpreted as a dust scattering halo \citep{kargaltsev12short}, while an asymmetrical structure which extended up to 2.5$'$ from the point source has been attributed to a \emph{magnetar wind nebula} \citep{younes12}.

Here we report on  three new \cxo\ observations that, complementing the \cxo, \swift, \xte\ and \xmm\ data already analysed by \citet{kargaltsev12short} and \citet{younes12}, allowed us to characterise better the long-term spectral and temporal behaviour of \src\ and  investigate in depth the nature of the diffuse emission. We also present the properties of the bursts observed with the Swift/BAT instrument, and the results of the search of the new SGR at radio frequencies using the 101-m-diameter Robert C. Byrd Green Bank Telescope (GBT).

\section{X-ray observations and data reduction}

\begin{table*}
\begin{minipage}{11.5cm}
\centering \caption{X-ray observations used for this work. \label{obs-log}}
\begin{tabular}{@{}lcccc}
\hline
Instrument$^a$ & Obs.ID & \multicolumn{2}{c}{Start / end time (UT)} & Exposure\\
 & & \multicolumn{2}{c}{(YYYY-MM-DD hh-mm-ss)} & (ks)\\
\hline
\cxo/ACIS-S$^b$ & 10126 & 2009-06-06 05:00:21 & 2009-06-06 18:42:35 & 46.6\\
\swift/XRT (PC) & 00458907000 & 2011-08-07 20:03:28 & 2011-08-07 21:33:36 & 2.2 \\
\swift/XRT (WT) & 00458907001 & 2011-08-07 21:45:01 & 2011-08-07 21:46:50 & 0.1 \\
\swift/XRT (WT) & 00458907002 & 2011-08-08 05:04:48 & 2011-08-08 05:31:59 & 0.2 \\
\swift/XRT (WT) & 00458907003 & 2011-08-08 14:46:07 & 2011-08-08 15:14:00 & 1.7 \\
\swift/XRT (WT) & 00458907004 & 2011-08-08 19:59:00 & 2011-08-08 20:15:29  & 1.0 \\
\swift/XRT (WT) & 00458907006 & 2011-08-09 00:44:54 & 2011-08-09 03:54:00 & 2.7 \\
\swift/XRT (WT) & 00458907007 & 2011-08-12 05:40:44 & 2011-08-12 10:45:00 & 4.9 \\
\swift/XRT (WT) & 00458907008 & 2011-08-14 04:13:00 & 2011-08-14 09:11:00 & 5.4 \\
\xte/PCA & 96434-01-03-00 & 2011-08-14 20:22:10 & 2011-08-14 21:56:10 & 6.9 \\
\swift/XRT (WT) & 00458907009 & 2011-08-18 02:55:55 & 2011-08-18 19:27:00& 8.1 \\
\xte/PCA & 96434-01-03-01 & 2011-08-18 17:25:56 & 2011-08-18 19:57:10 & 6.9 \\
\swift/XRT (WT) & 00458907010 & 2011-08-21 02:59:58 & 2011-08-21 17:42:00 & 2.5 \\
\cxo/ACIS-S & 14329 & 2011-08-22 15:34:26 & 2011-08-22 19:55:27 & 13.0 \\
\xte/PCA & 96434-01-04-00 & 2011-08-24 14:26:50 & 2011-08-24 16:56:10 & 6.7 \\
\swift/XRT (WT) & 00458907011 & 2011-08-24 19:17:59 & 2011-08-24 19:34:00 & 1.0 \\
\swift/XRT (WT) & 00458907012 & 2011-08-27 06:44:59 & 2011-08-27 10:03:00 & 2.0 \\
\xte/PCA & 96434-01-05-00 & 2011-08-29 10:30:28 & 2011-08-29 12:52:10 & 6.2 \\
\swift/XRT (PC) & 00501752000 & 2011-08-30 00:29:01 & 2011-08-30 02:29:24 & 2.6 \\
\swift/XRT (WT) & 00458907013 & 2011-08-30 07:26:37 & 2011-08-30 10:42:00 & 2.2 \\
\swift/XRT (PC) & 00458907014 & 2011-09-02 00:54:30 & 2011-09-02 21:44:57 & 2.1 \\
\xte/PCA & 96434-01-06-00 & 2011-09-02 13:23:28 & 2011-09-02 15:35:10 & 5.7 \\
\swift/XRT (PC) & 00458907015 & 2011-09-05 02:37:14 & 2011-09-05 06:15:56 & 1.7 \\
\cxo/ACIS-S & 14055 & 2011-09-05 11:36:40 & 2011-09-05 16:59:09 & 16.3\\
\swift/XRT (PC) & 00458907016 & 2011-09-10 08:12:44 & 2011-09-10 13:09:30 & 2.0 \\
\swift/XRT (WT) & 00032097001 & 2011-09-15 00:11:07 & 2011-09-1513:21:00 & 9.1\\
\xmm/EPIC & 0679380201 & 2011-09-17 15:04:37 & 2011-09-17 23:05:39 & 28.6 \\
\swift/XRT (WT) & 00032097002 & 2011-09-18 00:39:58 & 2011-09-18 11:59:59 & 10.4\\
\swift/XRT (WT) & 00032097003 & 2011-09-21 05:41:57 & 2011-09-21 23:11:59 & 7.7\\
\swift/XRT (WT) & 00032097004 & 2011-09-24 00:58:02 & 2011-09-25 00:00:00 & 8.1\\
\cxo/ACIS-S & 14056 & 2011-10-02 13:02:21 & 2011-10-02 21:13:55 & 24.5\\
\cxo/ACIS-S & 14057 & 2011-11-12 08:30:51 & 2011-11-12 20:25:38 & 37.6\\
\hline
\end{tabular}
\begin{list}{}{}
\item[$^{a}$] \swift/XRT photon counting (PC) provides two dimensional imaging information and a 2.5073-s time resolution; in windowed timing (WT) mode only one-dimensional imaging is preserved, achieving a time resolution of 1.766 ms. 
\item[$^{b}$] Observation of HESS\,J1834--087, see \citet*{misanovic11} and
\citet{kargaltsev12short}.
\end{list}
\end{minipage}
\end{table*}

\subsection{\cxo}

The field of \src\  was serendipitously imaged by \cxo\ on 2009 June 06 during observations targeting the extended TeV source HESS\,J1834--087 (see \citealt*{misanovic11} and \citealt{kargaltsev12short} for more details). Four more \cxo\ observations dedicated to \src\ (the first of which has already been published in \citealt{kargaltsev12short}) were carried out; the first was performed 15 days after the SGR activation (see again \citealt{kargaltsev12short}) and the last one after about 97 days (Table~\ref{obs-log}).

All these observations were carried out with the Advanced CCD Imaging Spectrometer (ACIS; \citealt{garmire03})  in Very Faint timed-exposure imaging mode. The source was positioned in the back-illuminated ACIS-S3 CCD (the only CCD used in these observations) at the nominal target position, and we used a sub-array of 1/8 (frame time: 0.44104 s). New level 2 event files were generated using the \cxo\ Interactive Analysis of Observation software (\textsc{ciao}) version 4.4. We used the latest ACIS gain map, and applied the time-dependent gain and charge transfer inefficiency corrections. The data were then filtered for bad event grades. 

\subsection{\swift, \xte, and \xmm}
The first \swift/XRT observation started right after the \swift/BAT  detection of the first burst on 2011 August 07 and further follow-up observations were carried out  for a total of 78 ks (Table~\ref{obs-log}), covering the first 50 d after the beginning of the outburst. The XRT data were processed and filtered with standard procedures and criteria using \textsc{ftools} tasks in the \textsc{heasoft} software package (v.6.11) and the calibration files in the 2011-08-12 \textsc{caldb} release. Similarly, BAT mask-tagged light curves, images and spectra were created using the standard BAT analysis software within \textsc{ftools}.

An observing campaign of \src\ was carried out also with the \emph{Rossi X-ray Timing Explorer} (\xte). Here we report on the Proportional Counter Array (PCA; \citealt{jahoda06}) data collected within $\sim$30 d from the onset of the outburst (obs. ID: 96434; see Table~\ref{obs-log}); after this period, the source flux become too low for the PCA sensitivity. We restricted our analysis to the data in Good Xenon mode, with a time resolution of 1 $\mu$s and 256 energy bins. The event-mode data were extracted in the 2--10 keV energy range from all active PCA units (in a given observation) and all layers, and binned into light curves of 0.1 s resolution using the standard \xte\ analysis software within \textsc{ftools}.

\xmm\ observed the field of \src\ twice: on 2005 September 18 and on 2011 September 17, about 40 days after the discovery of the source (Table~\ref{obs-log}). In the first observation ($\sim$20 ks) the EPIC pn \citep{struder01short} and MOS \citep{turner01short} detectors were operated in Full Frame mode. In the 2011 pointing ($\sim$24 ks), the pn detector was operating in Full Frame mode and the two MOSs in Small Window mode (see \citealt{younes12} for more details). The data were processed, screened and  checked for episodes of high particle background in a standard way using the \textsc{sas} software package version 11.0.

\section{Results}

\subsection{Timing analysis}

The arrival times of the 2--10 keV photons from \src\ were converted to  the barycenter of the Solar System. Pulsations at $\sim$2.48 s were clearly detected in all the \swift\ (WT), \xmm\ and \cxo\ data sets, and in the \xte\ ones up to $\sim$30 d after the onset of the outburst.

\begin{figure}
\resizebox{\hsize}{!}{\includegraphics[angle=0]{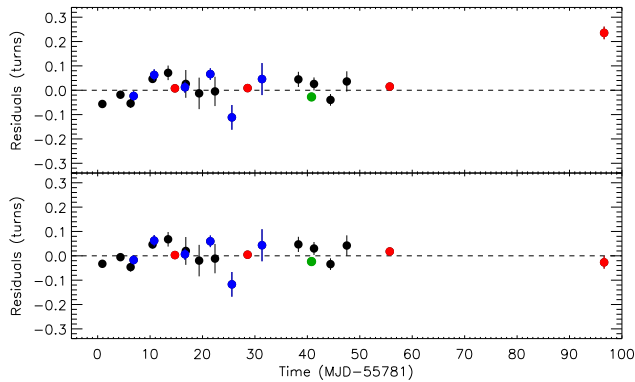}}
\caption{\label{phasefit} Best-fit residuals obtained using ephemeris A (top) and  ephemeris B (bottom). Black: \swift/XRT, blue: \xte/PCA, red: \cxo/ACIS-S, green: \xmm/EPIC. Note that the last \cxo\ point of the upper panel (day $\sim$97) has not been included in the fit; its residual is shown with respect to the extrapolation of the ephemeris A. (See the electronic journal for a colour version of this figure.)}
\end{figure}

The relative phases and amplitudes were such that, using a quadratic  function (i.e. including a $\dot{P}$ component), the signal phase evolution could be followed unambiguously for the observations from 2011 August 07 to 2011 October 02. The resulting fully-coherent solution (ephemeris `A', see Table~\ref{timing-fit} and Fig.~\ref{phasefit}) had a best-fit period of $P=2.482\,299\,90(4)$ s and $\dot{P}=8.28(2)\times10^{-12}$ s s$^{-1}$ (uncertainties here are 1$\sigma$); MJD 55781.0 was used as reference epoch. These values are consistent with the measurement reported by  \citet{kuiper11} and \citet{kargaltsev12short}, who used in their fitting data with a shorter time span (about two weeks and one month, respectively).

The inclusion of the last \cxo\ dataset (Obs. ID 14057, 2011 November 12) shows a significant disagreement ($\sim$9$\sigma$) with the value expected by extrapolating ephemeris A; a much better solution (the Fisher-test chance probability is $3.8\times10^{-4}$) can be obtained by including in the fit a higher-order (cubic) term, corresponding to a $\ddot{P}$ component. This results in period and period derivative values which are only slightly different from those of ephemeris A, $P=2.482\,300\,17(7)$ s and $\dot{P}=8.06(4)\times10^{-12}$ s s$^{-1}$ (see Fig.~\ref{phasefit} and Table~\ref{timing-fit} for the complete ephemeris, which we designate as `B').
\begin{table}
\centering \caption{Spin ephemeris of \src. We also give for convenience the corresponding period $P$ and period derivative $\dot{P}$, as well as the derived characteristic age $\tau_c=P/(2\dot{P})$, dipolar magnetic field $B\approx(3c^3IP\dot P/8\pi^2R^6)^{1/2}$, and rotational energy loss $\dot{E}=4\pi^2 I \dot{P}P^{-3}$ (here we took $R=10$ km and $I=10^{45}\ {\rm g\, cm}^2$ for the star radius and moment of inertia, respectively).}
\label{timing-fit}
\begin{tabular}{@{}lcc}
\hline
Parameter & \multicolumn{2}{c}{Value}\\
 & Ephemeris A$^a$ & Ephemeris B \\
\hline
 Range (MJD) & 55781.9--55836.7 & 55781.9--55877.6\\
 Epoch (MJD) &  \multicolumn{2}{c}{55781.0}\\
$\nu$ (Hz) & 0.402\,852\,210(7) & 0.402\,852\,17(1)\\
$\dot{\nu}$ (Hz s$^{-1}$) & $-1.344(3)\times10^{-12}$ & $-1.308(7)\times10^{-12}$\\
$\ddot{\nu}$ (Hz s$^{-2}$) & -- & $-1.2(3)\times10^{-20}$\\
\hline
\multicolumn{3}{c}{Derived parameters}\\
\hline
$P$ (s) &  2.482\,299\,90(4) & 2.482\,300\,17(7)\\
$\dot{P}$ (s s$^{-1}$) & $8.28(2)\times10^{-12}$ &  $8.06(4)\times10^{-12}$\\
$\ddot{P}$ (s s$^{-2}$) & --  & $7(2)\times10^{-20}$\\
$\tau_c$ (kyr) &  4.8 & 4.9\\
$B$ (G) &$ 1.5\times10^{14}$ & $1.4\times10^{14}$\\
$\dot{E}$ (\lum) & $2.1\times10^{34}$ & $2.1\times10^{34}$\\
\hline
\end{tabular}
\begin{list}{}{}
\item[$^{a}$] Fully-coherent timing solution.
\end{list}
\end{table}

\begin{figure}
\resizebox{\hsize}{!}{\includegraphics[angle=0]{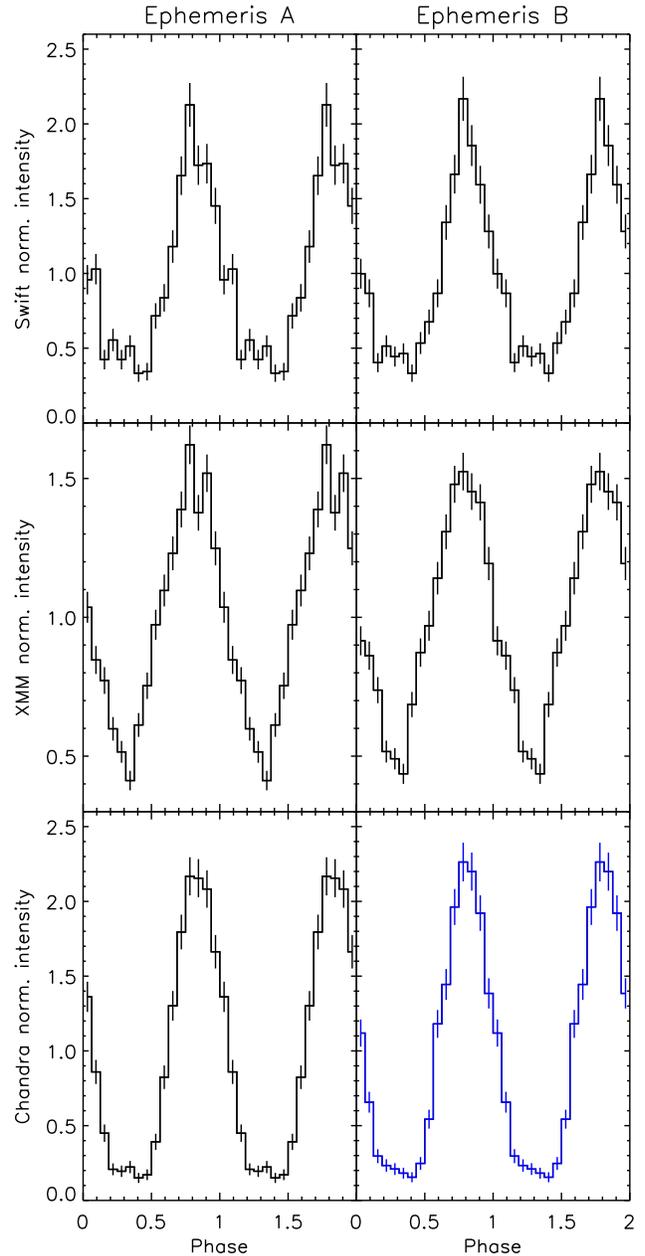}}
\caption{\label{efold} 16-bin \cxo, \xmm\ and \swift\ epoch-folded light curves (background-subtracted). The \cxo\ light curve obtained using ephemeris B (blue line, bottom right corner) includes also observation 14057.}
\end{figure}

In Fig.~\ref{efold} we show the \swift, \xmm\ and \cxo\ background-subtracted light curves (2--10 keV) obtained folding the data on both ephemeris A and B. As can be seen from the plot, the profiles obtained with the different ephemeris are all very similar. To quantify any phase shifts, for each instrument we cross-correlated the A and B pulse profiles and the resulting peak in the cross-correlation function was fit with a Gaussian. This yielded small phase lag values of $\Delta\phi=0.020\pm0.008$ cycles between the \swift\ A and B profiles, $0.02\pm0.01$ cycles between the \xmm\ A and B profiles, and of $0.031\pm0.004$ cycles between the \cxo\ A and B profiles. The test confirms that indeed ephemeris B introduces only a mild correction with respect to ephemeris A. The pulsed fraction, estimated by fitting a sinusoidal curve to the data, was $(93\pm4)\%$, $(88\pm4)\%$, $>$94\% ($1\sigma$ lower limit), and $(80\pm16)\%$ in the four \cxo\ observations in chronological order.

\subsection{Burst analysis}

\begin{figure}
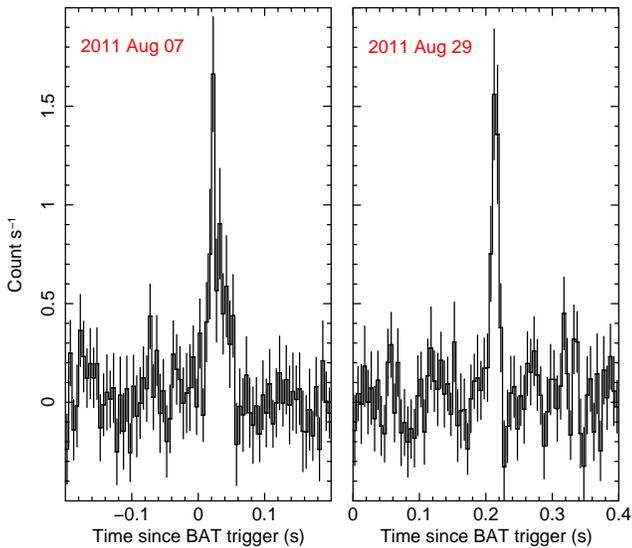

\resizebox{\hsize}{!}{\includegraphics[angle=-90]{burst_7000.eps}
\hspace{0.3cm}\includegraphics[angle=-90]{burst_2000.eps}}
\caption{\label{bursts} \swift/BAT 15--150 keV mask-weighted light curve of the bursts observed from \src\ (bin size: 5 ms).}
\end{figure}

We performed temporal and spectral analyses on the two triggered BAT events (Fig.~\ref{bursts}). For the first event (trigger 458907, 2011 August 07 at 19:57:46.36 UT) the $T_{90}$ duration (the time during which 90\% of the burst counts were accumulated) was 35 ms while the estimated total duration was 39 ms; for the second trigger (trigger 501752, 2011 August 29 at 23:41:12.04 UT) we found $T_{90}=11$ ms and a total duration of 12 ms. The values were computed by the \textsc{battblocks} task (based on a Bayesian blocks algorithm; \citealt{scargle98}) from 15--150 keV mask-weighted light curves with 1 ms bin size.

We fit the time-averaged spectra with simple functions: power law, blackbody and optically-thin thermal bremsstrahlung. They all provided statistically acceptable fits, but a blackbody model with $kT\sim10$ keV yielded lower $\chi^2$ values in both cases (Table~\ref{burst-spectra}). For this model, the $T_{90}$ fluences of the bursts in the 15--150 keV band were $\sim$$10^{-8}$ erg cm$^{-2}$  (first event) and $7\times10^{-9}$ erg cm$^{-2}$ (second event).

\begin{table*}
\begin{minipage}{9.5cm}
\centering
\caption{Spectral analysis results for the bursts (\swift/BAT data).}
\label{burst-spectra}
\begin{tabular}{@{}lccccc}
\hline
Trigger & Model & $kT$ & $\Gamma$ & Flux$^a$ & red. $\chi^2$ (dof) \\
 & & (keV) & & (\flux) \\
\hline
458907 & \textsc{pl} & -- & $3.1^{+0.2}_{-0.3}$ & $2.1\times10^{-7}$ & 1.22 (56) \\
 & \textsc{ottb} & $29^{+4}_{-5}$ & -- & $2.5\times10^{-7}$ & 1.00 (56) \\
 & \textsc{bb} & $9.3\pm0.8$ & -- & $2.8\times10^{-7}$ & 0.82 (56) \\
501752 & \textsc{pl} & -- & $2.4^{+0.4}_{-0.3}$ & $6.5\times10^{-7}$ & 0.72 (56) \\
 & \textsc{ottb} & $44^{+20}_{-12}$ & -- & $6.4\times10^{-7}$ & 0.70 (56) \\
 & \textsc{bb} & $11\pm1$ & -- & $6.1\times10^{-7}$ & 0.69 (56) \\
\hline
\end{tabular}
\begin{list}{}{}
\item[$^{a}$] In the 15--150 keV energy range.
\end{list}
\end{minipage}
\end{table*}

\subsection{Spectral analysis of the X-ray persistent emission}\label{xrayspectra}

For the spectroscopy (performed with the \textsc{xspec} 12.7 fitting package; \citealt{arnaud96}) we concentrate on the \cxo\ spectra which, owing to the ACIS-S high throughput, the long observing time and the superb \cxo\ point-spread function, are those with the best statistical quality and signal-to-noise ratio. We fit the spectra from the four 2011 observations simultaneously with the hydrogen column density tied between all data sets. Photons having energies below 2 keV and above 10 keV were ignored, owing to the very few counts from \src. The abundances used were those of \citet{anders89} and photoelectric absorption cross-sections were from \citet{balucinska92}.

Although the relatively small number of photons ($\sim$800 net counts for observations 14329 and 14055 each, $\sim$500 for observation 14056, and $\sim$50 for observation 14057) allowed us to carry out only a limited spectral analysis, there was clear evidence for a rather soft spectrum and high absorption toward the source. Both a steep power law (photon index $\Gamma\sim4$) and a blackbody with temperature $kT\sim1$ keV gave acceptable fits. The best-fit parameters for these spectral models are given in Table~\ref{cxo-fits}. No additional spectral components were statistically required. In general AXPs/SGRs, especially when in outburst, exhibit more complex (multi-component) spectra, usually fit by the superposition of two blackbodies, or a blackbody (or a resonant cyclotron scattering model; \citealt{rea08,zane09}) and a high-energy power law (e.g. \citealt{mte05short,esposito09short,bernardini09short,bernardini11short,icd07,rea09short}). However, similarly to the case of SGR\,1833--0832 \citep{esposito11short}, the relatively low flux and high absorption of \src\ made our \cxo\ observations not very sensitive to the presence of  additional spectral components.

\begin{table*}
\begin{minipage}{17cm}
\centering \caption{Spectral analysis of the \cxo\ observations. Errors are at a 1$\sigma$ confidence level.} \label{cxo-fits}
\begin{tabular}{@{}lccccccc}
\hline
Obs.ID & Model & \nh & $\Gamma$ & $kT$ & $R_{BB}$$^{a}$ & Flux (obs. / unabs.)$^b$ & $\chi^2_\nu$/dof\\
 & & ($10^{22}$ cm$^{-2}$)& & (keV) & (km) & ($10^{-12}$ \flux) & \\
\hline
\cxo/14329 & \multirow{4}{*}{\textsc{pl}} & \multirow{4}{*}{$19.9\pm1.5$} & $3.8\pm0.3$ & -- & -- & $1.9^{+0.1}_{-0.2}$ / $12\pm2$ & \multirow{4}{*}{0.77/117}\\
\cxo/14055 &  &  &  $4.2^{+0.3}_{-0.4}$ & -- & -- & $1.4\pm0.1$ / $10^{+3}_{-2}$ & \\
\cxo/14056 &  & & $4.6\pm0.4$ & -- & -- & $0.51\pm0.03$ /$5\pm1$  & \\
\cxo/14057 &  & & $6.1^{+1.8}_{-1.7}$ & -- & -- & $0.029_{-0.05}^{+0.06}$ /  $0.7_{-0.4}^{+1.1}$ & \\
\cline{1-2}
\cxo/14329 & \multirow{4}{*}{\textsc{bb}} & \multirow{4}{*}{$12.6^{+1.1}_{-1.0}$} & -- & $1.00^{+0.07}_{-0.06}$ & $0.37^{+0.07}_{-0.06}$ & $1.7\pm0.1$ / $4.2^{+0.4}_{-0.3}$ & \multirow{4}{*}{0.73/117} \\
\cxo/14055 &  &  & -- & $0.92\pm0.06$ & $0.41^{+0.08}_{-0.07}$ & $1.2\pm0.1$ / $3.4\pm0.3$ &\\
\cxo/14056 & & & -- & $0.83^{+0.05}_{-0.05}$ & $0.33^{+0.07}_{-0.05}$ & $0.47\pm0.03$ / $1.4^{+0.2}_{-0.1}$ & \\
\cxo/14057 & & & -- & $0.6^{+0.2}_{-0.1}$ & $0.2^{+0.3}_{-0.1}$ & $0.027\pm0.05$ / $0.12^{+0.09}_{-0.04}$ & \\
\hline
\end{tabular}
\begin{list}{}{}
\item[$^{a}$] The blackbody radius is calculated at infinity and for an arbitrary distance of 5.4 kpc.
\item[$^{b}$] In the 2--10 keV energy range.
\end{list}
\end{minipage}
\end{table*}

\begin{figure}
\resizebox{\hsize}{!}{\includegraphics[angle=-90]{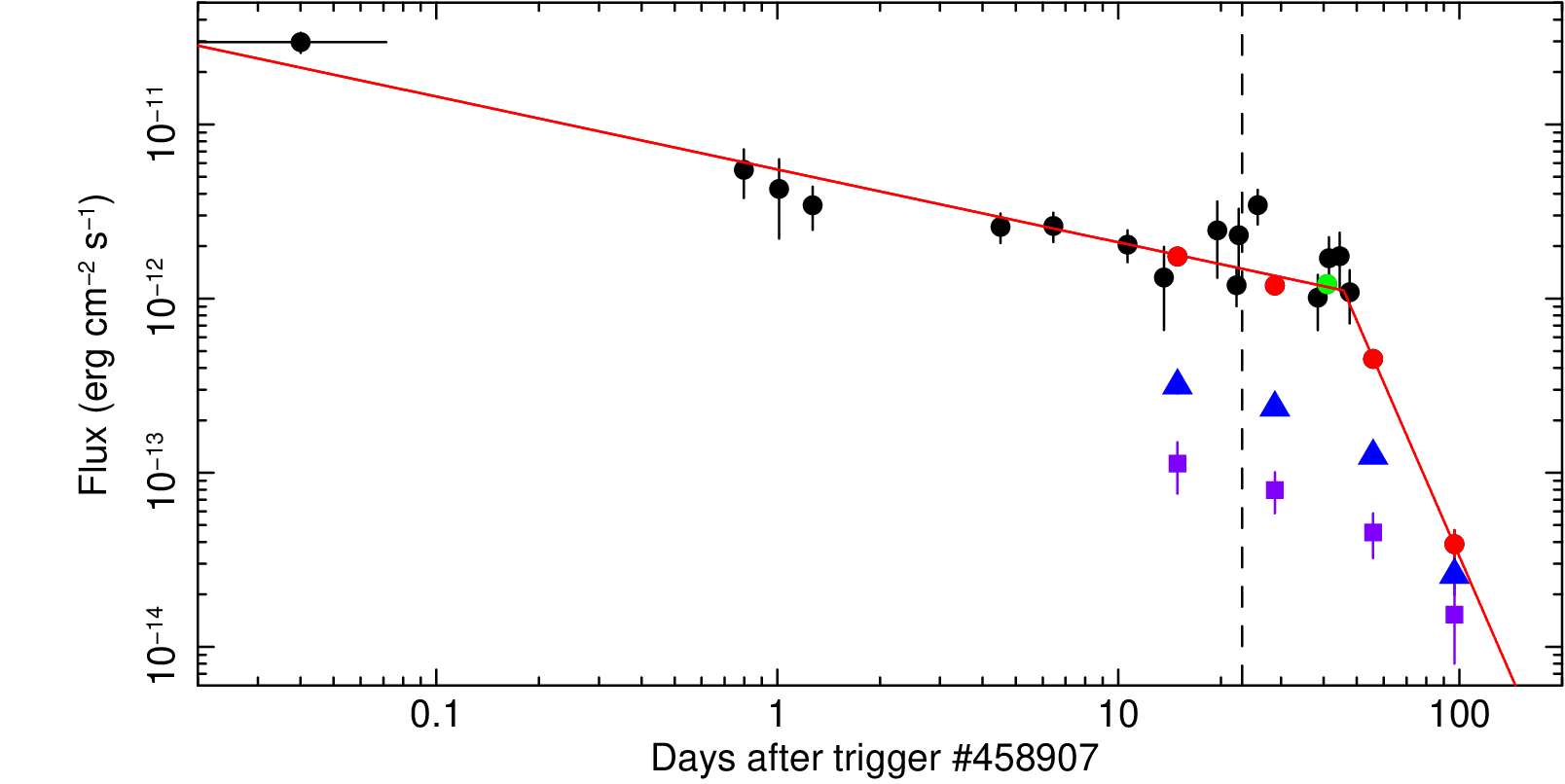}}
\caption{\label{decay}Time evolution of the (absorbed) flux of \src\ (filled circles) in the 2--10 keV energy range as measured
with \swift\ (black), \xmm\ (green) and \cxo\ (red). The 2--10 keV fluxes of the dust-scattering halo measured during the four
\cxo\ observations in the 2$''$--$10''$ region are shown by blue triangles, and those in the 10$''$--$20''$ region by
purple squares. We assumed as $t = 0$ the time of the first \swift/BAT trigger. The solid line represents the broken power law
model fit to the data (see Section~\ref{xrayspectra}). The vertical dashed line marks the epoch of the second BAT trigger
(\#501752). (See the electronic journal for a colour version of this figure.)} 
\end{figure}

During the time spanned by the first three \cxo\ observations, the spectrum softened and the X-ray flux decreased by $\sim $30\% in the first two weeks, and then by a further $\sim$60\%. To obtain flux measurements over the outburst, a blackbody model was fit to the \swift/XRT data simultaneously, with all parameters left free to vary except for the absorption column density, that was fixed at the \cxo\ value. This resulted in an acceptable fit ($\chi^2_{\nu}=0.91$ for 193 d.o.f.) with spectral parameters similar to those reported in Table~\ref{cxo-fits} but much more poorly constrained. We plot the resulting long-term light curve in Fig.~\ref{decay}.
The flux evolution could be satisfactorily described by a broken power law ($\chi^2_\nu=1.27$ for 16 dof, see  Fig.~\ref{decay}). Fixing $t = 0$ at the time of the first burst, the break occurred at $(46\pm1)$ d, when the index changed from $\alpha_1 = 0.42\pm0.02$ to $\alpha_2 = 4.5 \pm 0.4$; the flux at the break time is $(1.11\pm0.3)\times10^{-12}$ \flux. This result is in agreement with the trend reported in \cite{kargaltsev12short}, whose observations cover only the first part of the decay.

\subsection{Diffuse emission}\label{diffuse}

The diffuse emission reported by \citet{kargaltsev12short} around \src\ for observation 14329 was clearly detected also in the new \cxo\ observations. Considering the large absorption derived from the X-ray spectra, it was probably due to the scattering of the point source radiation by dust along the line of sight, as also observed in other magnetars \citep{tiengo10short,rivera10,esposito11short,olausen11}. The small field of view of the ACIS instrument used in 1/8 subarray mode (width of 1$'$) allowed us to probe the inner part of the diffuse emission only. The more extended structure which \citet{younes12} interpreted as a \emph{magnetar wind nebula} could not be investigated with these observations.

Owing to their longer path to the observer, the X-rays scattered by dust at a distance $d_{\rm dust}$ are detected with a time delay which increases with the off-axis angle $\theta$ (in arcsec) according to:
\begin{equation} \Delta t= 1.4\times 10^{-5} \frac{x d}{1-x} \theta^2~~{\rm days,} \label{delay} \end{equation} 
where $d$ is the distance of the X-ray source (in kpc) and \mbox{$x\equiv d_{\rm dust}/d$} \citep{truemper73}. As a consequence, the halo photons detected in our \cxo\ observations (within $\sim$30$''$ from the central source) had a short time delay, less than 1 day (unless most of the dust is concentrated extremely close to \src, say a few tens of pc). We then expect the ratio between the scattered and transmitted flux to be the same in all the \cxo\ observations.

For each \cxo\ observation, we extracted the spectra from three annuli around the point source, with radii of $2''$--$10''$, $10''$--$20''$, and $20''$--$30''$. The background spectrum was extracted from two $20''$-radius circles centred at more than $1'$ away from \src. Based on ChaRT/MARX\footnote{See http://cxc.harvard.edu/chart/index.html for details on the Chandra Ray Tracer (ChaRT) and Model of AXAF Response to X-rays (MARX) software packages.} simulations of the \cxo\ point-spread function (PSF), we expect $\sim$5\%, $\sim$1\%, and $\sim$0.5\% of the counts to come from the point-source in the $2''$--$10''$, $10''$--$20''$, and $20''$--$30''$ annuli, respectively. We estimated the halo flux by fitting the background-subtracted spectra with an absorbed power-law model to which we added a (fixed) blackbody component (parameters are those of the best fits in Table~\ref{cxo-fits}, with normalisations opportunely rescaled) to account for the small contamination from the point source estimated above. The resulting 2--10 keV fluxes of the two innermost annuli are plotted in Fig.~\ref{decay}; the spectra of the $20''$--$30''$ region, owing to their poor count statistics, do not provide useful information and are not considered in the following. As expected, the halo fluxes decreased with time. We performed the same analysis\footnote{In this case no contribution from the central source needed to be subtracted, since \src\ was not detected in this observation (see \citealt{kargaltsev12short} for a detailed analysis).} also on the \cxo\ pre-outburst observation (obs. ID 10126), obtaining 2--10 keV fluxes of $(1.0\pm0.3)\times10^{-14}$ \flux\ and  $(2.4\pm0.9)\times10^{-14}$  \flux\ for the diffuse emission in the $2''$--$10''$ and  $10''$--$20''$ regions, respectively.

\begin{figure}
\resizebox{\hsize}{!}{\includegraphics[angle=-90]{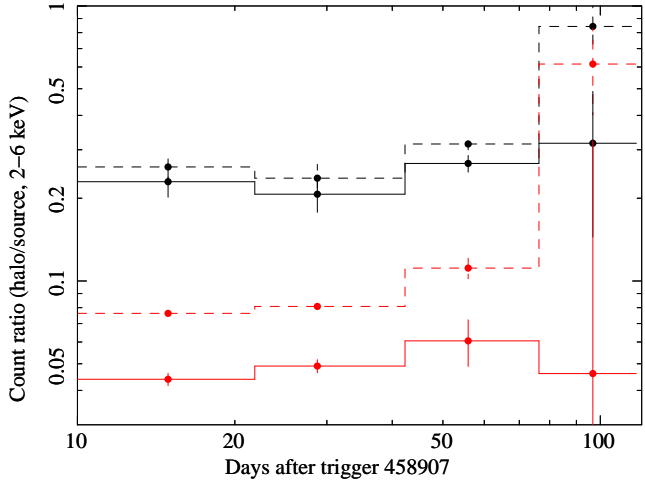}}
\caption{\label{ratios} Ratio of the halo and point-source count rates (black 2$''$--10$''$ region, red 10$''$--20$''$ region) in the four post-outburst \cxo\ observations (dashed lines). Solid lines indicate the corresponding ratio obtained after the subtraction of the count rates detected during the pre-outburst \cxo\ observation. A significant increase with time is observed only if the emission of the pre-outburst observation is not subtracted. (See the electronic journal for a colour version of this figure.)} 
\end{figure}

Since the observations were all carried out with the same detector and set-up, we could study the evolution in time of the relative contribution of the diffuse emission against the point source intensity by directly comparing the count rates in different regions.
While the ratios between the net count rates of the halo in the two annuli and that of the point-source remained constant (within the uncertainties) during the first three \cxo\ post-outburst observations, they significantly increased in the last one. This suggests the presence of an additional contribution to the fluxes of the diffuse emission. Indeed, if we subtract from each of the above count rates the net count rate observed in the same regions during the pre-outburst observation, the count rate ratios are consistent with being constant in time (see Fig.~\ref{ratios}). This means that, if one assumes that the diffuse emission detected in the pre-outburst observation was present at the same level also in the most recent observations, the flux evolution of the diffuse emission detected after the outburst agrees with the expectations for a dust-scattering halo.

For a given model of scattered halo (consisting of a dust-scattering cross section and of the assumed properties of the grains), the fractional halo intensity (as a function of the angle $\theta$ and the photon energy) depends on the amount of the dust along the line of sight and its spatial distribution.  As discussed below, there is evidence that most of the dust in the direction of \src\ is concentrated in a few discrete clouds. We therefore consider for simplicity the case of a single cloud responsible for most (or all) of the observed scattered radiation. We use the dust cross section in the Rayleigh-Gans approximation and a power-law distribution for the dimensions of the grains, with power-law index $q=-3.5$ and maximum and minimum grain size $a_{\rm max}=0.25~\mu{\rm m}$ and $a_{\rm min}=0.005~\mu{\rm m}$ \citep{mathis77}. Based on Eq.\,\ref{delay}, we can neglect the time delay at the small angular radii we are considering.

\begin{figure}
\resizebox{\hsize}{!}
{\includegraphics[angle=0]{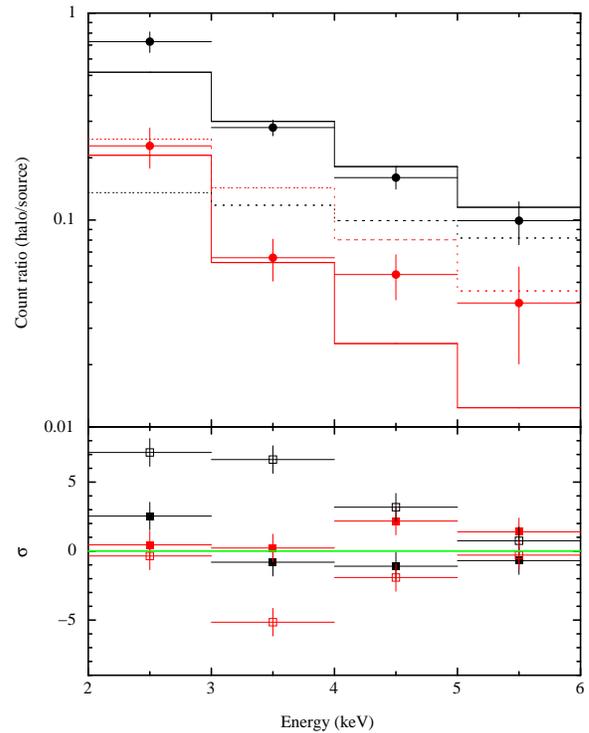}}
\caption{\label{cxoratios} Ratio of the halo and point-source count rates in the sum of the first three post-outburst \cxo\ observations (black $2''$--$10''$ region, red $10''$--$20''$ region). The best-fit dust-scattering halo model and its residuals (in standard-deviation units) are indicated by the solid lines and filled-squares, while the dotted lines and open squares correspond to the model and residuals obtained by fixing $x=d_{\rm dust}/d$ = 0.9. Assuming a distance of 5 kpc for \src, the distance between the dust and the neutron star would be $185\pm20$ pc in our best-fit model and 500 pc for  $x=0.9$. (See the electronic journal for a colour version of this figure.)} 
\end{figure}

To compare the predictions of this simple model with the data, we merged the first three post-outburst observations, which have the largest signal-to-noise ratio, and computed the halo-to-point-source ratios in the $2''$--$10''$ and $10''$--$20''$ regions and four energy bands in the 2--6 keV range. As done for the individual observations, the background and the diffuse emission in the pre-outburst observation were subtracted from the count rates of the different regions and the count ratios were corrected to account for the instrumental PSF. The observed ratios were fit with our halo model ($\chi^2_\nu=2.62$ for 6 dof, see  Fig.~\ref{cxoratios}), obtaining  $x=d_{\rm dust}/d=0.963\pm0.004$ and a normalization corresponding to \nh\ $=(7.1\pm0.4)\times10^{22}$ cm$^{-2}$  in the dust cloud, which is a significant fraction of the total hydrogen column density derived from the X-ray spectral analysis. Although an optimization of the dust model is beyond the scope of this work, we have checked that these results are not substantially affected by varying the dust-grain parameters within a reasonable range. In particular, we can robustly conclude that the dust cloud must be necessarily close to \src\ to generate the compact X-ray halo we observed with \cxo. In fact, as the $x$ parameter decreases, the halo counts in the 10$''$--$20''$ region progressively increase with respect to those in the inner region, indicating that the halo becomes significantly broader (as an example, see the best-fit obtained by fixing $x$ = 0.9 in Fig.~\ref{cxoratios}).

\section{Radio observations with the Green Bank Telescope}

Radio observations of \src\ were taken at the Green Bank Radio telescope on 2011 August 18, 11 days after the first X-ray burst, and on three other occasions separated by $\sim$1 month one from the other. Data were acquired at a central frequency of 2.0 GHz over a bandwidth of 800 MHz split into 512-MHz frequency channels, and 8-bit sampled every 655 $\mu$s. The duration of the first three observations (performed on August 18, September 20 and October 19) was 23 minutes while the last (on November 22) lasted 32 minutes.

About 100 MHz of the total bandwidth were affected by strong narrow-frequency radio interferences hence the interested frequency channels have not been considered in the analysis. After de-dispersing the signal over a wide range of dispersion measure (DM) values (0--3000 pc cm$^{-3}$), data were analysed both blindly, searching for significant peaks on the power spectra of the fast Fourier transformed time series, for single pulses on the time series directly, and by folding the de-dispersed data at the period obtained from X-ray observations. No signal was detected above a flux density of 0.05 mJy, which is the limit derived from the radiometer equation (see e.g. \citealt{manchester01short}) for a signal-to-noise ratio of 10. This rather high ratio has been chosen for caution because the data (especially the last ones) were affected also by long-period, wide-band radio frequency interferences.

Also the three pointings of the Parkes Multibeam Pulsar Survey (e.g. \citealt{manchester01short}) closest to the source position were analysed in search of a pulsed (periodic or sporadic) signal. Nothing was found in any of the archival observations dating 1998 December 1, 1999 July 14, and 2003 October 18, down to a flux density limit of 0.22 mJy for the 1999 pointing (at 5$'$ from the source) and 0.32 mJy for the two other pontings (at 7$'$ from the source).

\section{Discussion}

In this work we reported on the long-term evolution of the magnetar \src\ and its X-ray halo during the 2011 outburst using all available data from \cxo, \xmm, \swift, and \xte.  In particular, we presented three new observations obtained with the \cxo/ACIS-S instrument. The last observation was carried out on 2011 November 12, about 100 days after the onset of the bursting activity that had led to the discovery of the source. This long time span enabled us to refine the rotational ephemeris available for the source. The extended monitoring of \src\ also revealed a monotonic decrease of the flux. The rate of change of the source luminosity varied in time. After a first phase during which the flux decreased following a rather shallow power-law ($F\propto t^{-0.42}$ up to $\sim$46 d from the onset), the decay became much faster, with $F\propto t^{-4.5}$. The trend we observed prior to the break is compatible with that reported in \cite{kargaltsev12short}, who, however, could not detect the change in slope because of the shorter time coverage of their observations.

The flux decay patterns observed in outbursting magnetars are quite diverse but can be grouped into three main classes: exponential, power-law and broken power-law \cite[see e.g.][]{rea11}. In this respect, the behaviour of \src\ bears some resemblance to that of the low-field magnetar, SGR\,0418+5729 \cite[][]{esposito10short}. In both sources, the flux decay followed a broken power-law, which became steeper at later times. For SGR\,0418+5729 the analysis of the flux evolution, together with that of the spectral parameters (blackbody temperature and radius) and of the pulse profiles, made it possible to derive information on the size and thermal evolution of the hot spot which is believed to have formed on the star surface as a consequence of the outburst. In particular, observations seem to favour a picture in which a hotter zone appeared within (or close to) one of the two warm polar regions \cite[][]{esposito10short}. While the temperature of the hot spot decreased following the same power-law at all times, its radius was roughly constant before the break and shrunk afterwards.

\begin{figure}
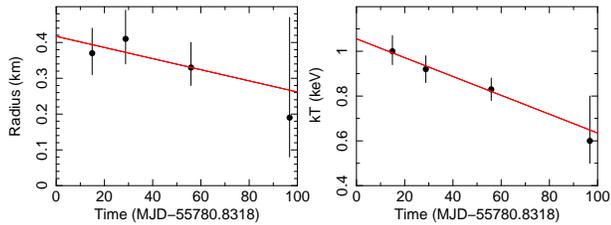

\resizebox{\hsize}{!}{\includegraphics[angle=-90]{radius.eps}\hspace{-1.5cm}\includegraphics[angle=-90]{kt.eps}}
\caption{\label{discuss}Time evolution of the blackbody parameters of \src\ inferred from the spectral analysis of the \cxo\ data. The radius is evaluated at infinity and assuming a distance of 5.4 kpc. A linear fit is superimposed on both data sets.}
\end{figure}

Although data for \src\ are much scantier, the four \cxo\ observations have enough statistics to attempt a similar analysis. The time evolution of the blackbody temperature and radius, as derived from the spectral fit of \cxo\ data, is shown in Fig.~\ref{discuss}.  Here the quite large errors prevent to reach any firm conclusion; still, it is interesting to note that also in the case of \src\ the data are consistent with a picture in which the radius stays constant (or slightly increases) before the break and then decreases, while the temperature drop is well represented by a monotonic decrease. Therefore also in this source the break in the flux evolution could be related to the sudden change in the way the radius evolves, the temperature playing little role in this respect.

Observations of the decay phase in magnetars can provide valuable information on the physical process(es) at the basis of the energy released during the outburst. Up to now two main possibilities have been discussed: energy injection deep in the crust, possibly produced by magnetic dissipation \cite[e.g.][]{let02,ponsrea12}, and surface heating induced by the returning currents which flow along the closed field lines of the twisted magnetosphere \cite[e.g.][]{tlk02,belo09}. More data are needed before a firm conclusion can be drawn. Here we just mention that, according to more recent simulations of the evolution of the surface temperature in a magnetar following an energy deposition in the crust (\citealt{ponsrea12}; see also Yakovlev et al., in preparation\footnote{See http://www.rikkyo.ne.jp/web/z5000063/magnetar/8yakovlev.pdf}), the radius of the heated region is fairly constant in time and only increases slightly at late times, while the temperature monotonically decreases. Even the latest 2D heat transfer codes have been, so far,Ê unsuccessful in explaining the decrease in size as observed in SGR\,0418+5729 and possibly also in \src.

The discovery of pulsed radio emission from a few transient magnetars (namely XTE\,J1810--197, 1E\,1547--5408, and PSR\,1622--4950; \citealt{camilo06,camilo07,levin10short,anderson12short}), gave support to the possibility that magnetars and ordinary radio pulsars are linked at some level (see \citealt{perna11} and \citealt{pons11} for theoretical work supporting this connection with respect to the ourburst phenomenology). Indeed, \cite{rea12} recently suggested that magnetars radio emission, despite having peculiar characteristics, might be powered by the same physical mechanism responsible for the radio emission in ordinary pulsars. In particular, particle acceleration in a voltage gap, followed by a pair cascade, could be present also in the magnetosphere of a magnetar, the apparently different properties of the radio emission in the two classes of sources being possibly related to the twisted magnetic field of the  magnetars. In this environment the pair cascade might in fact proceed differently, the radio emission be more variable and unstable, and the radio spectrum be flatter due to the larger electron density that a twisted magnetosphere can sustain with respect to an ordinary pulsar. 

Interestingly, the magnetars that showed radio pulsed emission so far were found to have rotational energy loss rates larger than their X-ray luminosity during quiescence \citep{rea12}. Assuming a distance of 5.4 kpc, the quiescent luminosity of \src\ is $L_X <1.7\times10^{31}$ \lum\ (which corresponds to the flux limit derived from the archival \cxo\ observation by \citealt{kargaltsev12short}). The rotational power is $\sim$$2.1\times10^{34}$ \lum\ and the ratio $L_{\rm X}/L_{\rm rot}\sim 8\times10^{-4}$ is well below unity. This makes \src\ similar to the other radio emitting magnetars, hence it should be expected to show radio emission if the scenario by \citet{rea12} is correct. However, the non-detection of pulsed radio emission from \src\ could be due to unfavourable beaming, or, given the intermittent nature of the magnetar radio emission (e.g. \citealt{burgay09}), \src\ was simply not active at the time of our observations. We also note that, while our present limit for \src\ (1.5 mJy kpc$^2$ for $d=5.4$ kpc) is substantially lower than the observed radio luminosity of the known radio magnetars ($\sim$100--400 mJy kpc$^2$ at their peak; \citealt{camilo06,camilo07,levin10short}) it may be, however, non entirely constraining, since more than one hundred ordinary radio pulsars have luminositiy below this value.\footnote{See the online version of the Australia Telescope National Facility (ATNF) pulsar catalogue \citep{manchester05} at http://www.atnf.csiro.au/research/pulsar/psrcat/.} We thus cannot exclude that \src\ was active in radio but far less luminous than the known radio magnetars and too faint to be detected.

\subsection{Diffuse emission} 
The four \cxo\ observations taken during the 2011 outburst decay of \src\ showed a significant decrease in the flux of the diffuse X-ray emission extending up to 20$''$ from the neutron star. This variability strongly supports the dust-scattering halo origin already suggested by \citet{kargaltsev12short}. At such small angles, the ratio of the halo and point-source count rates are expected to be the same in all the observations, since the time delay of the halo photons is short enough to make the intrinsic variation of the source flux negligible [see Eq.\,(\ref{delay}) and Fig.~\ref{decay}]. The data are consistent with such time-independent ratios, provided that the extended emission observed in 2009 \citep{kargaltsev12short} was present also during the 2011 outburst. We compared the energy and spatial dependence of the fractional halo intensity with the predictions of a simple dust scattering model, in which the dust has a power-law grain-size distribution and is concentrated in a single cloud. The narrow radial profile of the halo requires the dust cloud to be close to the SGR, at a fractional distance $x\approx0.96$ (see Fig.~\ref{cxoratios}).

\begin{figure}
\resizebox{\hsize}{!}{\includegraphics[angle=0]{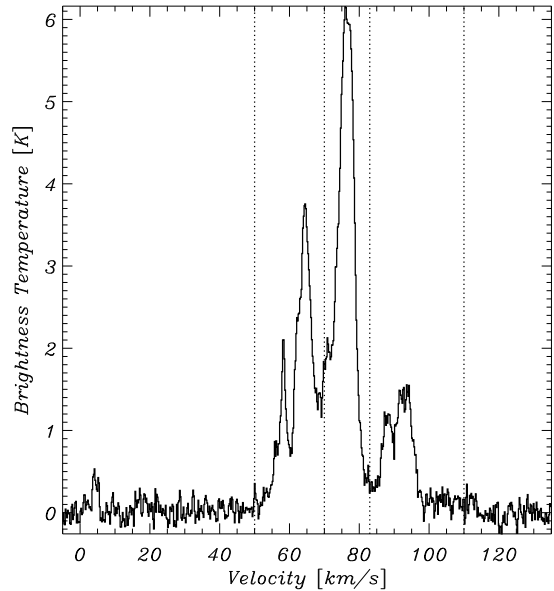}}
\caption{\label{CO} CO spectrum extracted from a circular region with radius 30 arcsec around \src. The dotted lines indicates
the three velocity intervals corresponding to the images of figure \ref{COmaps} (50--70, 70--83, and \mbox{83--110} km s$^{-1}$).}
\end{figure}

The dust distribution along the line of sight can be estimated from radio observations of the $^{13}$CO (J = 1$\rightarrow$0) line (110.2 GHz), since this molecule is a good tracer of interstellar dust. We retrieved the data in the direction of \src\ from the high-spatial-resolution Boston University--Five College Radio Astronomical Observatory (BU--FCRAO) Galactic Ring Survey obtained with the FCRAO radio telescope \citep{jackson06short}. The velocity-resolved line profile is shown in Fig.~\ref{CO}. No strong emission at radial velocities below 50 km s$^{-1}$, which for this direction correspond to distances $<$4 kpc, is visible. This excludes the presence of a significant amount of CO in the Local and Sagittarius spiral arms. The line profile shows prominent peaks at velocities compatible with the Scutum, Norma, and 3-kpc arms. The highest peak in the line profile indicates that the largest CO concentration along this line of sight, using the Galactic rotational curve described in \citet*{sofue09}, is located at $d\sim5.2$ kpc in the Norma arm. Assuming that this dust cloud was the responsible for the scattering halo, the source would be located at a distance of $\sim$5.4 kpc.

\begin{figure*}
\resizebox{\hsize}{!}{\includegraphics[angle=0]{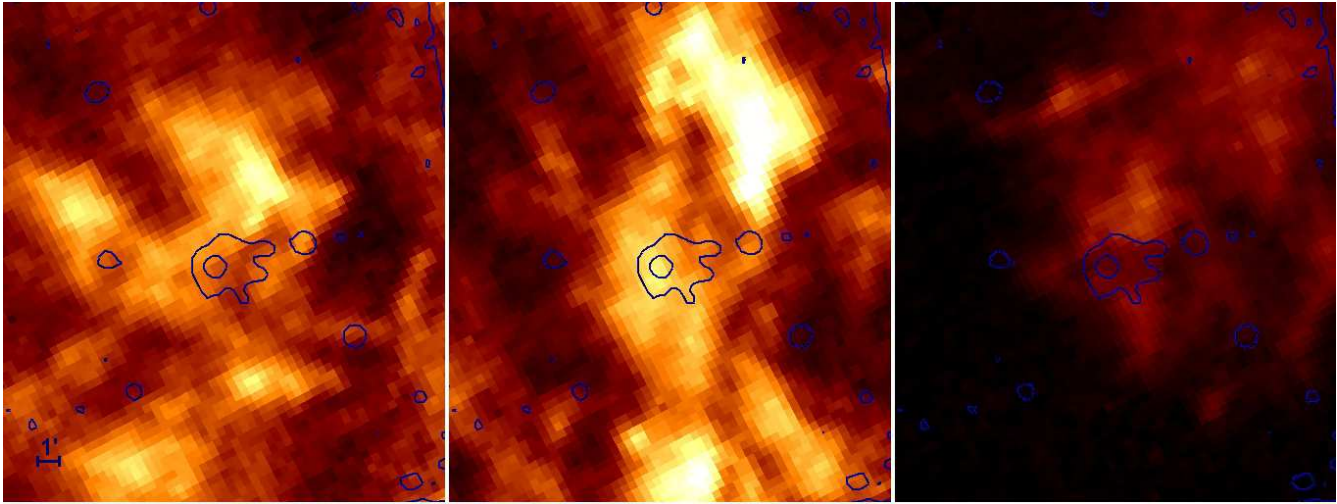}}
\caption{\label{COmaps} CO distribution in the region around \src\ in celestial coordinates. The three maps were obtained integrating the intervals of velocities shown in Fig.~\ref{CO}. The contours represent the X-rays intensity in the band 2--4 keV from the \xmm\ EPIC-pn post-outburst observation. (See the electronic journal for a colour version of this figure.)}
\end{figure*}

Although the sky coverage of the 2011 \cxo\ data was too small to include the X-ray diffuse emission extending to $\sim$$2'$--$3'$ observed with \xmm, our analysis provides some evidence that also this feature could be due to scattering by the same dust cloud responsible for the compact X-ray halo detected by \cxo. This interpretation would easily explain also the possible variability seen between the two \xmm\ observations about 6 yr apart \cite[][]{younes12}. If confirmed by further observations, the variability of an extended feature  $\sim$10 light years across (for $d\sim5$ kpc) would be difficult to explain in the wind nebula interpretation, but compatible with a projection effect in a dust-scattering halo. The spatial asymmetry of the emission, which extended to the SW, can be explained by a non-uniform dust distribution in the plane of the sky. The CO sky maps of this region (Fig.~\ref{COmaps}), obtained selecting the CO data in the three velocity intervals shown in figure \ref{CO}, show indeed a patchy structure. Although no clear correlation is visible between the CO maps and the X-ray surface brightness, the presence of large discontinuities on angular scales $<$$1'$ indicates an inhomogeneous dust distribution in clouds, over a spatial scale quite compatible with what implied by the observed asymmetry of the X-ray halo.

At variance with what we found for the emission within 20$''$ from the source, at the arcmin scale the ratio of the diffuse to point-source emission had different values in the two \xmm\ observations: the flux of \src\ in 2011 was $\sim$30 times larger than in 2005, while the extended emission was brighter by only a factor $\sim$2 (see table 1 in \citealt{younes12}). This is not a problem in the dust scattering interpretation because, for the distances we derived above ($d_{\mathrm{dust}}\sim5.2$ kpc, $d\sim5.4$ kpc), angular distances between 50$''$ and $\sim$200$''$ correspond to an average time delay of $\sim$1 month [see Eq.\,(\ref{delay})]. Therefore, the halo photons detected in the 2011 \xmm\ observation were emitted in the first days of the outburst (which started $\sim$40 days earlier), when the source was brighter. As suggested by \citet{younes12} based on the non-detection of \src\ in the deep \cxo\ observation carried out in 2009, this SGR might have experienced an outburst shortly before the 2005 \xmm\ observation. In the lack of direct observations of such a putative outburst, we assume it had the same flux evolution of the 2011 outburst (Fig.~\ref{decay}). Since the flux of \src\ in the 2005 observation \citep{younes12} is similar to that measured in our last \cxo\ observation and the previous \cxo\ observation was executed $\sim$40 days before, which approximately corresponds to the time delay expected for the photons of the large scale diffuse emission, these photons were emitted when the source was $\sim$15 times brighter. Therefore, this significantly larger flux can explain the different ratio between the halo and the point source flux observed in 2005 and in 2011. Note that such an explanation requires that the scattering dust is located close to the source, otherwise the time delay would be too small to account for the required difference in the fluxes. In conclusion, both the small and large-scale X-ray halos can be explained only if they are produced by dust close to \src.

\section*{Acknowledgments} This research is based on data and software provided by the \cxo\ X-ray Center (CXC, operated for NASA by SAO under contract NAS8-03060), the NASA/GSFC's HEASARC, and the ESA's \xmm\ SOC. This publication makes use of molecular line data from the BU-FCRAO Galactic Ring Survey, a project funded by the NSF.  We also used data collected with the GBT operated by the NRAO, which is a facility of the NSF operated under cooperative agreement by Associated Universities, Inc.  We thank the anonymous referee for valuable comments. NR is supported by a Ram\'on y Cajal fellowship. This work was partially supported by ASI through ASI--INAF contracts I/009/10/0, I/004/11/0, by INAF through grant PRIN-INAF 2010, and by NASA through \cxo\ Award Number G02-13076X issued by the CXC.

\bibliographystyle{mn2e}
\bibliography{biblio}

\bsp

\label{lastpage}

\end{document}